\documentclass[conference]{IEEEtran}
 \IEEEoverridecommandlockouts

\usepackage[utf8]{inputenc} 
\usepackage[T1]{fontenc}
\usepackage{url}
\usepackage{ifthen}
\usepackage[cmex10]{amsmath} 
\usepackage[pdftex]{graphicx}
\usepackage{amssymb,amsfonts}
\usepackage{mathtools}
\usepackage[cmintegrals]{newtxmath}

\newtheorem{lemma} {Lemma}

\usepackage{cite}
\usepackage{xcolor}
\usepackage{authblk}

\newtheorem{thm}{Theorem}

\newtheorem{cor}{Corollary}

\usepackage[top=1in,bottom=1.1in,left=0.65in,right=0.65in]{geometry}

\begin{document}
\title{Characterization of Deletion/Substitution Channel Capacity for Small Deletion and Substitution Probabilities}

\author[*]{Mohammad Kazemi\thanks{M. Kazemi's work was funded by UK Research and Innovation (UKRI) under the UK government’s Horizon Europe funding guarantee [grant number 101103430]. T. M. Duman's work was funded by the European Union ERC TRANCIDS 101054904.
Views and opinions expressed are however those of the author(s) only and do not necessarily reflect those of the European Union or the European Research Council Executive Agency. Neither the European Union nor the granting authority can be held responsible for them. 
}}
\author[**]{Tolga M. Duman}
\affil[*]{Dept. of Electrical and Electronic Engineering, Imperial College London}
\affil[**]{Dept. of Electrical and Electronics Engineering, Bilkent University
\protect\\Email:  mohammad.kazemi@imperial.ac.uk, duman@ee.bilkent.edu.tr}

\maketitle

\begin{abstract}
We consider binary input deletion/substitution channels, which model certain types of synchronization errors encountered in practice. Specifically, we focus on the regime of small deletion and substitution probabilities, and by extending an approach developed for the deletion-only channel, we obtain an asymptotic characterization of the channel capacity for independent and identically distributed (i.i.d.) deletion/substitution channels. To do so, given a target probability of successful decoding, we first develop an upper bound on the codebook size for arbitrary but fixed numbers of deletions and substitutions, and then extend the result to the case of random deletions and substitutions to obtain a bound on the channel capacity. Our final result is: The i.i.d. deletion/substitution channel capacity is approximately \(1 - H(p_d) - H(p_s)\), for \(p_d, p_s \approx0\), where \(p_d\) and \(p_s\) are the deletion and substitution probabilities, respectively.
\end{abstract}

\section{Introduction}
Channels with insertions, deletions, and substitutions are crucial in modeling certain communication systems with synchronization errors. Such errors are observed in cases like clock mismatches in wireless communication systems, as well as in bit-patterned media recording and DNA data storage channels.
In this paper, our focus is on the special case of channels experiencing bit losses (deletions) and bit errors (substitutions). Specifically, we are interested in independent and identically distributed (i.i.d.) deletions and substitutions at the regime of small deletion and substitution probabilities. Our objective is to characterize the channel capacity for this asymptotic regime.

Dobrushin established the information stability of channels with memoryless synchronization errors in \cite{Dobrushin67}, which is extended in \cite{Morozov24} to synchronization errors governed by a Markov chain. However, obtaining the exact capacity has proved very difficult. Even for a channel with only i.i.d. deletions, the capacity remains unknown. As a result, the literature has focused on obtaining bounds on the capacity.
Gallager's work on sequential decoding for channels with synchronization errors provides the first lower bound on the capacity of a deletion channel \cite{Gallager1961}, which states that for deletion probabilities of less than $1/2$, i.e., $p_d<1/2$, the capacity of i.i.d. binary deletion channel (BDC) is not smaller than that of a binary symmetric channel with a cross-over probability of $p_d$; that is, the i.i.d. deletion channel capacity, denoted by $C_d$, satisfies
\begin{equation}
C_d \ge 1-H(p_d),
\end{equation} 
where $H(\cdot)$ denotes the binary entropy function. Indeed, Gallager's work is more general and establishes a lower bound on the capacity of i.i.d. insertion, deletion, and substitution error channels. For the case of deletion/substitution channels (i.e., with insertion probability of zero), the lower bound on the capacity, denoted by $C_{d,s}$, becomes:
\begin{equation}
C_{d,s} \ge 1 + p_d \log p_d + p_s \log p_s + p_c \log p_c, 
\end{equation}
\noindent where $p_s$ is the substitution probability, and $p_c = 1-p_d-p_s$ is the probability of correct transmission. In the regime of asymptotically small synchronization errors, the lower bound is approximately $1 - H(p_d) - H(p_s)$.

We will prove that this bound is asymptotically tight by establishing an upper bound on the capacity whose dominant terms match with it in the asymptotic regime of $p_d,p_s \rightarrow 0$.
There are also other attempts to obtain lower bounds on the capacity of channels with synchronization errors. The authors in \cite{Zigangirov69} employ tree codes to obtain lower bounds for insertion/deletion channels. 
In \cite{Mitzenmacher2006}, 
a lower bound on the BDC capacity is obtained using the jigsaw-puzzle decoding approach, which is further improved in \cite{Drinea2007}. 

In \cite{Diggavi2007}, the authors provide two upper bounds on the BDC capacity: a numerically evaluated one for any arbitrary deletion probability and a simple closed-form one. In \cite{Fertonani2010}, numerical upper bounds are provided for deletion/substitution channels by exploiting an auxiliary system where the receiver is provided with side information. 
The authors extend their results to a non-binary deletion channel in \cite{Rahmati13}, and to more general synchronization error channels with insertions, deletions, and substitutions in \cite{Fertonani10}. The authors of \cite{Rubinstein2023} improve the BDC lower bound in \cite{Drinea2007} as well as the upper bound in \cite{Fertonani2010}. In \cite{Cheraghchi19}, by converting the capacity problem into maximization of a real-valued, often concave, univariate function, the authors obtain upper bounds for a general class of repetition channels.

Several works have focused on approximating the capacity of synchronization error channels for the asymptotic regime of small synchronization error probabilities. 
In \cite{Kanoria10,Kanoria13}, the BDC capacity is approximated using series expansions up to the second and third leading terms, respectively.
Similar results are obtained for duplication channels in \cite{Ramezani13} and insertion channels in \cite{Tegin2024}.

More in line with our work, the authors of \cite{Kalai2010} show that, for small deletion probabilities, the i.i.d. BDC capacity can be upper-bounded as 
\begin{equation}
C \le 1-(1-o(1))H(p_d).
\label{Kalai}
\end{equation} 
To obtain this bound, they first obtain lower and upper bounds on the success probability of a guesser that, given the received erroneous sequence, outputs the transmitted sequence along with the deletion pattern. To this end, assuming that the decoding is successful with a certain probability, they obtain bounds on the number of deletion patterns that can be mistaken for the actual one at the receiver.
Finally, combining \eqref{Kalai} with Gallager's lower bound, they show that the BDC capacity approaches $1-H(p_d)$ as $p_d \rightarrow 0$. 
The authors extend the result to the $k$-copy BDC in \cite{Haeupler2014}.

In this paper, we adapt the general methodology in \cite{Kalai2010}, and extend the results to channels with substitution errors as well as deletions. Given a probability of successful decoding, we first provide an upper bound on the codebook size for deletion/substitution channels.
To this end, we first fix the numbers of deletions and substitutions and assume that the receiver knows these numbers as side information (it knows the number of deletions anyway due to the received sequence length), and obtain an upper bound on the codebook size, which is also an upper bound on the codebook size of the case without side information since side information cannot decrease the number of possible codewords. 
Finally, we extend the result to the case of random numbers of deletions and substitutions, and, combining with the result in \cite{Gallager1961}, we obtain a tight approximation for the channel capacity in the asymptotic regime of small deletion and substitution probabilities by letting the codeword length go to infinity.  

Throughout the paper, $|X|$ indicates the cardinality of the set $X$, and $\land$ and $\lor$ denote logical conjunction and disjunction, respectively. The discrepancy set between two sequences of equal lengths, $A$ and $B$, denoted by $\Delta(A,B)$ is defined as the set of indices in which $A$ and $B$ are in disagreements, i.e.,
$\Delta(A,B) = \{i | a_i \neq b_i\}$ with $a_i$ and $b_i$ being the $i$-th element of $A$ and $B$, respectively. Then, the distance between two sequences is defined as the size of their discrepancy set, which is equal to their Hamming distance. Viewing sequences as sets, their symmetric difference is defined as $\delta(A,B)= (A \backslash B) \cup (B \backslash A)$, where $\backslash$ indicates the set difference. In addition, $Pr_{x\in U^S}[T(x)]$ denotes the probability of $T$ holding, over $x$ chosen uniformly at random from the set $S$.

\section{System Model}
We consider a deletion/substitution channel, where each bit is independently deleted with probability $p_d$, flipped (substituted) with probability $p_s$, or correctly transmitted with probability $1-p_d-p_s$. 
We consider a codebook $C \subseteq \{0, 1\}^n$ consisting of messages of $n$ bits and of size $N = |C|$. We may think of a transmission pattern $A_t=a^t_1, a^t_2, . . . , a^t_{n-q_d}$ as a strictly increasing subsequence of $[n] := \{1, 2, . . . , n\}$, representing which bits are transmitted (not deleted), where $q_d$ is the number of deletions. 
Similarly, we denote a substitution pattern $A_s$ as a strictly increasing subsequence of $[n-q_d]$, $A_s = a^s_1, a^s_2, . . . , a^s_{q_s}$, where $q_s$ is the number of substitutions.

\begin{figure}
	\centering
	\includegraphics[scale = 0.6]{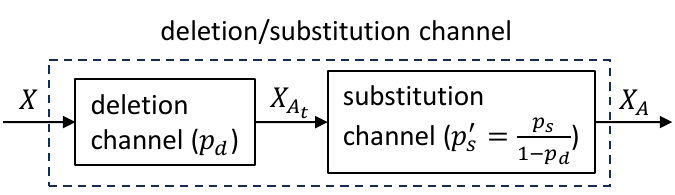}
	\caption{Decomposition of a deletion/substitution channel with deletion probability $p_d$ and substitution probability $p_s$ as a concatenation of a deletion channel with deletion probability $p_d$ followed by a substitution channel with substitution probability $p_s^{\prime}=\frac{p_s}{1-p_d}$.}
	\label{SysMod}
\end{figure}

A deletion/substitution channel with deletion probability $p_d$ and substitution probability $p_s$ can be decomposed as a concatenation of a deletion channel with deletion probability $p_d$ followed by a substitution channel with substitution probability $p_s^{\prime}=\frac{p_s}{1-p_d}$, as depicted in Fig. \ref{SysMod}, which we found to be easier to work with for our analytical derivations.

Each realization of the deletion/insertion channel can be represented as a pattern pair $A=(A_t,A_s)$.
For a string $X \in \{0, 1\}^n$, $X_{A}$ represents the transmission of $X$ through a deletion/substitution channel with transmission and substitution patterns $A_t$ and $A_s$, respectively, in the obvious way: the bits of $X$ are deleted according to the transmission pattern $A_t$, then the remaining bits are flipped with the substitution pattern $A_s$.

\section{Fixing the Numbers of Deletions and Substitutions}
We aim to obtain an upper bound on the size of a uniquely decodable codebook, given a probability of successful decoding.
We first obtain the upper bound on the codebook size for fixed numbers of deletions and substitutions, assuming the receiver knows these values as side information. Note that the receiver is always aware of the number of deletions, as it is equal to the length loss of the received sequence, but not the number of substitutions. 

To obtain the upper bound on the codebook size, we first obtain upper and lower bounds on the success probability of a guesser that, given a received sequence, guesses the transmitted codeword as well as the transmission (deletion) and substitution patterns. Then, we compare these upper and lower bounds to obtain an upper bound on the codebook size. 

In the following, we provide an upper bound on the success probability of a general guesser with side information. 

\begin{lemma}[Taken from \cite{Kalai2010}] 
Let $\rho$ be a joint distribution over $S\times T$, for
finite sets $S$ and $T$, such that the marginal distribution over $S$ is
uniform. Let $g: T\rightarrow S$ be a function. 
Then, for 
$(a,b)\sim \rho$,
$Pr[g(b)=a]\le\frac{|T|}{|S|}$.
\label{lemma0}
\end{lemma}

\begin{cor} Using Lemma \ref{lemma0}, for fixed numbers of deletions and substitutions, we obtain the following upper bound on the success probability of a general guesser:
\begin{equation}
    Pr_{Z \in U^{\mathcal{C}}} [g(Z_A)=(Z,A)] \le \frac{2^{n-q_d}}{N {n \choose q_d} {n-q_d \choose q_s}},
\end{equation}
\label{corollary1}
\end{cor}
since the possible number of transmitted codewords $Z$ is equal to the codebook size $N$ and the number of possible deletion/substitution patterns is ${n \choose q_d} {n-q_d \choose q_s}$. Also, since the output length is equal to $n-q_d$, the number of possible output sequences is upper bounded by $2^{n-q_d}$.

To obtain a lower bound on the success probability, we first need to obtain an upper bound on the number of neighboring patterns that can be mistaken for the actual one. 
To this end, we first upper bound the probability of two transmission/substitution pattern pairs leading to the same output for all possible inputs in the following lemma by extending the result in \cite{Kalai2010} to the case of both deletion and substitution errors.

\begin{lemma}
 Take any two transmission/substitution pattern pairs, $A=(A_t,A_s)$ and
$B=(B_t,B_s)$. For uniformly random $X \in \{0, 1\}^n$, $Pr_{X \in U^{\{0,1\}^n}} [X_A = X_B | A,B] \le 2^{-|\Delta_t|}$ if $\delta_s \subseteq \Delta_t$, and zero otherwise,
where $\Delta_t$ denotes the discrepancy set between the two transmission patterns and $\delta_s$ is the symmetric difference of the two substitution patterns; i.e., $\Delta_t=\Delta(A_t,B_t)$ and $\delta_s=\delta(A_s,B_s)$. 
Note that $\Delta_t$ is defined over the indices of the elements of the two sequences, not on their values.
	\label{lemma1}
\end{lemma}

\begin{IEEEproof} In \cite{Kalai2010}, it is shown that in the case of only deletions, the probability of channel outputs being the same is upper bounded as $Pr_{X \in U^{\{0,1\}^n}} [X_{A_t} = X_{B_t}] \le 2^{-|\Delta_t|}$. In the case of only substitutions, two outputs can be the same only if the two substitution patterns are exactly the same. However, that is not the case for the deletion/substitution scenario. This is because, as long as the substitution pattern difference set is a subset of the discrepancy set of the transmission patterns, the two outputs can be the same with the same probability as the two transmission patterns that lead to the same output; i.e., $Pr[X_A=X_B|\delta_s \subseteq  \Delta_t]=Pr[X_{A_t}=X_{B_t}]$. We obtain
\begin{equation}
\begin{split}
	Pr_{X \in U^{\{0,1\}^n}} &[X_A = X_B | A,B] \\
	&\begin{cases}
		=1,& \text{if } A=B,\\
		=0,& \text{if } A_t=B_t , A_s\neq B_s, \\
		\le 2^{-|\Delta_t|},& \text{if } A_t\neq B_t , A_s= B_s,\\
		=0,& \text{if } A_t\neq B_t , A_s\neq B_s , \delta_s \not \subseteq \Delta_t,\\
		\le 2^{-|\Delta_t|},& \text{if } A_t\neq B_t , A_s\neq B_s , \delta_s \subseteq \Delta_t.\\
	\end{cases}    
\end{split}
\end{equation}
Simplifying this gives us the final expression in the lemma.
\end{IEEEproof}

\begin{cor} The expression in Lemma \ref{lemma1} can be further upper bounded as $Pr_{X \in U^\{0,1\}^n} [X_A = X_B | A,B] \le 2^{-|\Delta_t|}, \forall \delta_s$.
\label{corollary3}
\end{cor}

In the following lemma, we upper bound the number of patterns in a certain neighborhood of a given pattern that can be mistaken for it.

\begin{lemma} For any pattern $A=(A_t,A_s)$ and integers $s$ and $t$, the number of patterns $B=(B_t,B_s)$ with $|\Delta_t| \le t$ and $|\delta_s| \le s$ that can be mistaken with $A$ is upper bounded by
\begin{equation} 
(s+1) (t+1) {q_s +s \choose q_s}  {t + s \choose s} {{q_d+t}\choose {q_d}} {{2q_d+t+1}\choose{2q_d+1}} .
\label{lem2}
\end{equation}
\label{lemma2}
\end{lemma}
	
\begin{IEEEproof} From Lemma \ref{lemma1}, two patterns can only be mistaken with each other if the set of their substitution differences is a subset of their transmission pattern differences; i.e., $\delta_s \subseteq \Delta_t$. The number of such patterns, $I$, can be expressed as
\begin{equation}
\begin{split}
    &I = |\{B | \delta_s \subseteq \Delta_t, |\Delta_t|\le t, |\delta_s|\le s\}|\\
    &=\! \sum_{t^\prime = 0}^t \underbrace{|\{B_t | |\Delta_t|=t^\prime\}|}_{I_{t^\prime}}
    \!\times\! \underbrace{|\{B_s | \delta_s \!\subseteq \Delta_t, |\Delta_t|=t^\prime, |\delta_s|\!\le s\}|}_{I_s} .
\end{split}
\end{equation}
$I_{t^\prime}$ is upper bounded in \cite{Kalai2010} as	$I_{t^\prime} \le {{2q_d+t^\prime+1}\choose{2q_d+1}} {{q_d+t^\prime}\choose {q_d}}$.

For $I_s$, we have 
\begin{equation}
\begin{split}
    I_s &=  |\{B_s | \delta_s \subseteq \Delta_t, |\Delta_t|=t^\prime, |\delta_s|\le s\}|\\
    &= \sum_{s^\prime=0}^{s}  |\{B_s | \delta_s \subseteq \Delta_t, |\Delta_t|=t^\prime, |\delta_s|= s^\prime\}|\\
    &\le \sum_{s^\prime=0}^{s} {q_s \choose q_s-s^\prime} {t^\prime \choose s^\prime},
\end{split}
\end{equation}
which, since ${q_s \choose q_s-s^\prime} \le {q_s +s^\prime \choose q_s} \le {q_s +s \choose q_s}$,
and ${t^\prime \choose s^\prime} \le {t^\prime + s^\prime \choose s^\prime} \le {t^\prime + s \choose s}$,
can be upper bounded as
\begin{equation}
\begin{split}
    I_s \le (s+1) {q_s +s \choose q_s} {t^\prime + s \choose s}.
\end{split}
\end{equation}

Summing up, we have
\begin{equation}
\begin{split}
    I \le& (s+1) {q_s +s \choose q_s} \sum_{t^\prime = 0}^t {{2q_d+t^\prime+1}\choose{2q_d+1}} {{q_d+t^\prime}\choose {q_d}} {t^\prime + s \choose s}\\
    \le& (s+1) (t+1) {q_s +s \choose q_s}  {t + s \choose s} {{q_d+t}\choose {q_d}} {{2q_d+t+1}\choose{2q_d+1}}  .
\end{split}
\end{equation}
\end{IEEEproof}

{\bf Definition:} For integers $t$ and $s$, we say that $X\in\{0,1\}^n$ is {\it $(t,s)$-bad} if there exist two patterns $A$ and $B$ such that $|\Delta_t|\ge t$, $|\delta_s|\ge s$, and $X_A=X_B$.

In the following lemma, we provide an upper bound on the number of $(t,s)$-bad sequences.

\begin{lemma} Fixing the numbers of deletions and substitutions, the number of $(t,s)$-bad sequences in $X\in \{0,1\}^n$, $I_{(t,s)-bad}$, can be upper bounded as
\begin{equation}
I_{(t,s)-bad} \le {n \choose q_d}^2 {n-q_d \choose q_s}^2 2^{n-t}.
\end{equation}
\end{lemma}

\begin{IEEEproof} 
We first obtain the probability that a randomly picked sequence is $(t,s)$-bad, denoted by $P_{(t,s)-bad}$. According to Corollary \ref{corollary3}, the probability that two patterns $A$ and $B$ lead to the same output is upper bounded by $2^{-|\Delta_t|}$, which is decreasing in $|\Delta_t|$. Then, using the union bound trick, we have
\begin{equation}
\begin{split}
       P_{(t,s)-bad}
   & \le \sum_{A,B s.t. |\Delta_t|\ge t } Pr_{X \in U^{\{0,1\}^n}} [X_A = X_B]\\
    &\le {n \choose q_d}^2 {n-q_d \choose q_s}^2 2^{-t}.
\end{split}
\end{equation}

Noting that there are a total of $2^n$ sequences, the upper bound on the number of $(t,s)$-bad sequences follows.
\end{IEEEproof}

\begin{thm}
Fixing the numbers of deletions and substitutions, suppose that there exists a decoding algorithm that succeeds with probability at least $\delta > 0$. Then, the size of the codebook satisfies
\begin{equation} 
    \log N \le n-q_d -\log{n \choose q_d} -\log{n-q_d \choose q_s} + \log \frac{2}{\delta} + \log \alpha,
    \label{theorem1}
\end{equation}
where 
$\alpha = t^2 \left(2 e\right)^{t-1} \left(\frac{5t}{q_s}\right)^{q_s}  \left(\frac{5t}{q_d}\right)^{3q_d+1}$,
with 
$t= \lceil{3 q_d} \log\frac{n e}{q_d}+ {3 q_s} \log\frac{(n-q_d) e}{q_s} + \log\frac{2}{\delta} \rceil.$
\end{thm}

\begin{IEEEproof} We first assume that the receiver knows the numbers of deletions and substitutions as side information. We consider a guesser that, given the received sequence $Z_A$, outputs the transmitted sequence $Z$ and the pattern pair $A=(A_t,A_s)$, both chosen uniformly at random, with a non-negligible probability. More specifically, the guesser first decodes the input $Z_A$ with the proposed decoder $R(\cdot)$ with success probability $\delta$, then outputs $g(Z_A)=(R(Z_A),A^\prime)$, where $A^\prime$ is the lexicographically first sequence to satisfy  $(R(Z_A))_{A^\prime}=Z_A$ if one exists, and the fixed pattern $A^\prime=(A_t^\prime,A_s^\prime)$, with $A_t^\prime=1,2,\cdots,n-q_d$ and $A_s^\prime=1,2,...,q_s$, otherwise.

We now lower bound the success probability of the guesser. For this, we first consider the probability that the decoding succeeds ($R(Z_A)=Z$) and the codeword $Z \in \mathcal{C}$ is not $(t,s)$-bad as follows
\begin{align}
    &Pr_{Z\in U^\mathcal{C}}[R(Z_A)=Z \wedge Z \text{ is not (t,s)-bad}] \nonumber\\
	&\ge Pr_{Z \in U^\mathcal{C}}[R(Z_A)=Z] + Pr_{Z \in U^\mathcal{C}}[ Z \text{ is not {(t,s)-bad}}] -1 \nonumber\\
	&= \delta - Pr_{Z \in U^\mathcal{C}}[ Z \text{ is {(t,s)-bad}}] \nonumber\\
	&\ge \delta - \frac{2^{n-t}}{N} {n \choose q_d}^2 {n-q_d \choose q_s}^2  \label{delta21}
\end{align}
since $Pr_{Z \in U^\mathcal{C}}[ Z \text{ is {(t,s)}-bad}] \le \frac{|\text{{(t,s)}-bad sequences} \in \{0,1\}^n |}{|\mathcal{C}|}$. 

If $N \le \frac{2^{n-q_d}}{{n \choose q_d} {n-q_d \choose q_s}}$, the theorem follows trivially. For $N \ge \frac{2^{n-q_d}}{{n \choose q_d} {n-q_d \choose q_s}}$, we have
\begin{equation}
\begin{split}
    \frac{2^{n-t}}{N} {n \choose q_d}^2 {n-q_d \choose q_s}^2 &\le 2^{q_d-t} {n \choose q_d}^3 {n-q_d \choose q_s}^3\\
    &\le 2^{q_d-t} \left(\frac{n e}{q_d}\right)^{3 q_d} \left(\frac{(n-q_d) e}{q_s}\right)^{3 q_s}.
\end{split}
\label{eq16}
\end{equation}
The right hand side is made smaller than or equal to $\frac{\delta}{2}$ if $t$ is selected as $t \ge q_d + {3 q_d} \log\frac{n e}{q_d}+ {3 q_s} \log\frac{e (n-q_d)}{q_s} + \log\frac{2}{\delta}$. We take $t= \lceil q_d +{3 q_d} \log\frac{n e}{q_d}+ {3 q_s} \log\frac{(n-q_d) e}{q_s} + \log\frac{2}{\delta} \rceil$, which lower bounds the right hand side of \eqref{eq16} by $\frac{\delta}{2}$.

If $Z$ is not $(t,s)$-bad, pattern $B$ gets mixed up with the desired pattern $A$ only when $\Delta_t \le t-1$ and $\delta_s \le s-1$. According to Lemma \ref{lemma2}, the number of such patterns is upper bounded by
\begin{align}
    & s t {q_s +s-1 \choose q_s}  {t + s-2 \choose s-1} {{q_d+t-1}\choose {q_d}} {{2q_d+t}\choose{2q_d+1}}\nonumber\\
    &\le  t^2 {q_s +t-1 \choose q_s}  {2t-2 \choose t-1} {{q_d+t-1}\choose {q_d}} {{2q_d+t}\choose{2q_d+1}}\nonumber\\
    &\le  t^2 \left(e \frac{q_s +t-1}{q_s}\right)^{q_s}  \left(2 e\right)^{t-1} \left(e\frac{q_d+t-1}{q_d}\right)^{q_d} \left(e\frac{2q_d+t}{2q_d+1}\right)^{2q_d+1}\nonumber\\
    &\le  t^2 \left(2 e\right)^{t-1} \left(\frac{5t}{q_s}\right)^{q_s}  \left(\frac{5t}{q_d}\right)^{3q_d+1}\nonumber\\
    &:=\alpha,
\end{align}
which is obtained by knowing that $s\le t$ due to $\delta_s \subseteq \Delta_t$ for two patterns to be mistaken with each other, and using the fact that ${n \choose k} \le \left(ne/k\right)^k$. We also used $t\ge 3 q_d$ and $t\ge 3 q_s$, since $t \ge q_d+{3 q_d} \log\frac{n e}{q_d}+ {3 q_s} \log\frac{e (n-q_d)}{q_s} + \log\frac{2}{\delta}$.

Conditioned on the decoding succeeding and the codeword not being $(t,s)$-bad, each deletion/substitution pattern is equally likely, and hence, the lexicographically first pattern is correct with probability at least $\alpha^{-1}$. Hence, the total success probability of the guesser is at least 	
\begin{equation}
		Pr_{Z \in U^{\mathcal{C}}} [g(Z_A)=(Z,A)] \ge \frac{\delta}{2} \alpha^{-1}.
\end{equation}

Finally, combining this lower bound with the upper bound in Corollary 1, we have 
\begin{align}
	\frac{\delta}{2} \alpha^{-1} &\le \frac{2^{n-q_d}}{N {n \choose q_d} {n-q_d \choose q_s}}
\end{align}
which gives us an upper bound on the codebook size for the case with side information at the receiver as

\begin{align}
	\log N  &\le n - q_d -\log { {n \choose q_d}} -\log { {n-q_d \choose q_s}} + \log \frac{2}{\delta} +\log \alpha.
\end{align}
Since side information cannot decrease the codebook size, this also serves as an upper bound on the codebook size for the case without side information. This concludes the proof.
\end{IEEEproof} 

\section{Capacity of I.I.D. Deletion/Substitution Channel}
Going from the case of fixed numbers of deletions and substitutions to the case where the numbers of deletions and substitutions are random merely involves taking advantage of the concentration of the numbers of deletions and substitutions around their means, $np_d$ and $n(1-p_d)p_s^{\prime}$, respectively. 

\begin{thm}
In the setup with random numbers of deletions and substitutions, suppose that there exists a decoding algorithm that succeeds with probability at least $\delta$ for codeword length of $n\ge \frac{12 \log \frac{8}{\delta}}{\min\{p_d,(1-p_d)p_s^{\prime}\}}$. Then, the codebook size $N$ satisfies
\begin{equation} 
\begin{split}
    \log N \le n - q_{d,\min} -\log{n \choose q_{d,\min}} &-\log{n-q_{d,\max} \choose q_{s,\min}} \\ 
    &+ \log \frac{4}{\delta}+ \log \alpha_{\max}^*,
\end{split}
\end{equation}
where $q_{d,\min}\!=\!(1\!-\!\gamma)np_d$, $q_{s,\min}\!=\!(1\!-\!\gamma)n(1\!-\!p_d)p_s^{\prime}$, $q_{d,\max}\!=\!(1\!+\!\gamma)np_d$, $q_{s,\max}\!=\!(1\!+\!\gamma)n(1\!-\!p_d)p_s^{\prime}$, with $\gamma\!=\!\sqrt{\frac{3 \log \frac{8}{\delta}}{np_d}}$, and
$\alpha_{\max}^* \!=\! {t_{\max}^{* 2}} \left(2 e\right)^{t_{\max}^*-1} \left(\frac{5t_{\max}^*}{q_{s,\max}^*}\right)^{q_{s,\max}^*}  \left(\frac{5t_{\max}^*}{q_{d,\max}^*}\right)^{3q_{d,\max}^*+1}$,
with
$t_{\max}^*=\lceil q_{d,\max}^*+{3 q_{d,\max}^*} \log\frac{n e}{q_{d,\max}^*}+ {3 q_{s,\max}^*} \log\frac{e (n-q_{d,\max}^*)}{q_{s,\max}^*} + \log\frac{4}{\delta} \rceil$.
\end{thm}

\begin{IEEEproof}
Suppose that we have a decoding algorithm for the deletion/substitution channel that succeeds on codebook $\mathcal{C}$ with probability $\delta > 0$. 
The standard multiplicative Chernoff bound gives an upper bound on the probability that a random variable lies outside an interval around its mean. More specifically (see \cite{Mitzenmacher2005}[Corollary 4.6]), if $X_i$'s are Bernoulli random variables with success probability $p_i$, then, 
$Pr(| \sum_i X_i -\mu|\ge \gamma \mu)\le 2 e^{-\frac{\mu \gamma^2}{3}}$, where $\mu=\sum_i p_i$. With $q_d=\sum_i X_i$, which is the random variable representing the number of deletions, we have $Pr(| q_d - np_d|\le \gamma np_d)\ge 1-2 e^{-\frac{np_d \gamma^2}{3}}$.
Setting the probability that the number of deletions, with mean $\mu_d=np_d$, lies in the interval $(1-\gamma)np_d \le q_d \le (1+\gamma)np_d$ to be at least $1-\frac{\delta}{4}$, we have $1-2 e^{-\frac{np_d \gamma^2}{3}}=1-\frac{\delta}{4}$. This gives us $\gamma=\sqrt{\frac{3 \log \frac{8}{\delta}}{np_d}}$. Finally, setting $\gamma\le\frac{1}{2}$, as a design choice, we get, $n\ge \frac{12 \log \frac{8}{\delta}}{np_d}$. 
Similarly, for the probability of the number of substitutions, with mean $\mu_s=n(1-p_d)p_s^{\prime}$, lying in the interval $(1-\gamma)n(1-p_d)p_s^{\prime} \le q_s \le (1+\gamma)n(1-p_d)p_s^{\prime}$ to be at least $1-\frac{\delta}{4}$, we need $n\ge \frac{12 \log \frac{8}{\delta}}{(1-p_d)p_s^{\prime}}$. Combining the two conditions, we finally have $n\ge \frac{12 \log \frac{8}{\delta}}{\min\{p_d,(1-p_d)p_s^{\prime}\}}$, as stated in the theorem. 
Then, we can write:
\begin{align}
        &Pr\{ \text{Successful decoding} \land |q_d-\mu_d| \!\le \!\gamma \mu_d \land |q_s-\mu_s| \!\le \!\gamma \mu_s\} \nonumber\\
        &\ge Pr\{ \text{Successful decoding}\} + Pr\{ |q_d-\mu_d|\le \gamma \mu_d \}\nonumber\\
        &\hspace{4.6cm} +Pr\{|q_s-\mu_s|\le \gamma \mu_s \} -2 \nonumber\\
        &\ge \frac{\delta}{2}.
\end{align}

Hence, there must be some $(q_d^*,q_s^*)$ in the defined intervals such that the success probability of the exact $(q_d^*,q_s^*)$ deletion/substitution channel is at least $\frac{\delta}{2}$.

Let $\alpha^*$ be given by
$\alpha^* = {t^*}^2 \left(2 e\right)^{t^*-1} \left(\frac{5t^*}{q_s^*}\right)^{q_s^*}  \left(\frac{5t^*}{q_d^*}\right)^{3q_d^*+1}$,
where 
    $t^*=\lceil q_d^*+{3 q_d^*} \log\frac{n e}{q_d^*}+ {3 q_s^*} \log\frac{e (n-q_d^*)}{q_s^*} + \log\frac{4}{\delta} \rceil$.

Since $\alpha^*$ is a non-decreasing function of both $q_d$  and $q_s$, it is maximized with their maximum values. Therefore, using the result in Theorem 1, by putting in each term the values in the intervals that maximize it, concludes the proof.
\end{IEEEproof}

\begin{thm}
The capacity of a deletion/substitution channel can be upper-bounded as 
$C \le 1 - (1 - o(1))(H(p_d)+H(p_s^{\prime}))$.
\end{thm}

\begin{IEEEproof}
Consider any fixed $p_d$ with $n$ going to infinity. We have $-n(1-\gamma)p_d \log((1-\gamma)p_d) \le \log{n \choose (1-\gamma)np_d} \le -n(1-\gamma)p_d \log(\frac{1-\gamma}{e}p_d)$. So $\log{n \choose (1-\gamma)np_d}=n(H(p_d)+o(1))$, as $\gamma$ is vanishing. 
Similarly, for any fixed $p_s$, as $p_d$ goes to zero, we have $p_s^{\prime}\rightarrow p_s$. So, we have $\log{n-(1+\gamma)np_d \choose (1-\gamma)n(1-p_d)p_s^{\prime}}=n(H(p_s)+o(1))$. 
Further, it can be shown that $\log \alpha_{\max}^*=\mathcal{O}(n(p_d+p_d) \log (\log(1/p_d)+\log(1/p_s^{\prime}))) = o(n(H(p_d)+H(p_s))$ as $p_d$ and $p_d$ (hence, also $p_s^{\prime}$) go to $0$. 
Therefore, the three terms $n - \log{n  \choose (1-\gamma)np_d} - \log{n-(1+\gamma)np_d \choose (1-\gamma)n(1-p_d)p_s^{\prime}}$ dominate the right-hand side of the equation; Dividing both sides by $n$, we obtain $\frac{1}{n}\log N \le 1 - (1 - o(1))(H(p_d)+H(p_s))$, which completes the proof. Note that the $o(1)$ term is understood as going to 0 in the limit as first $n$ goes to infinity and then as $p_d$ and $p_s$ go to $0$. 
\end{IEEEproof}

\begin{cor}
Combined with the lower bound result in \cite{Mitzenmacher2005}, it can be observed that the capacity of the deletion/substitution channel approaches $1-H(p_d)-H(p_s)$ for the probabilities of deletion $p_d$ and substitution $p_s$ near zero.
\end{cor}

\section{Conclusions}
In this paper, we characterized the capacity of binary channels with synchronization errors, where transmitted bits are randomly and independently deleted and flipped. To this end, we first obtained an upper bound on the codebook size for a given probability of successful decoding, assuming that the numbers of deletions and substitutions are known at the receiver. We then extended the results to the case of random numbers of deletions and substitutions. Finally, combining this with an existing lower bound, we showed that, in the regime of asymptotically small deletion and substitution probabilities, the rate loss is equal to the sum of the binary entropies of the deletion and substitution probabilities. As a future work, it would be of interest to extend this result to channels with insertion, deletion, and substitution errors.

\end{document}